\documentclass[aps,pra,showpacs,twocolumn]{revtex4-1}
\usepackage{amsfonts}
\usepackage{amssymb}
\usepackage{amsmath}
\usepackage{graphicx}
\usepackage{epsfig}
\usepackage{color}
\usepackage{amsmath,bm}
\usepackage{booktabs}
\usepackage[colorlinks=true, linkcolor=blue, citecolor=blue, urlcolor=blue]{hyperref}

\begin{document}

\title{Resonant Fields Inducing Energy Towers in Lieb Quantum Spin Lattice}
\author{J. Y. Liu-Sun}
\author{Z. Song}
\email{songtc@nankai.edu.cn}

\begin{abstract}
We study a ferromagnetic XXZ Heisenberg model on a Lieb lattice. A set of
exact eigenstates is constructed based on the restricted spectrum generating
algebra (RSGA) when a resonant staggered magnetic field is applied. These
states are identical to the eigenstates of a system of two coupled angular
momenta. Furthermore, we find that the RSGA can be applied to other
eigenstates of the Lieb lattice in an approximate manner. Numerical
simulations reveal that there exist sets of eigenstates, which obey a
quasi-RSGA. These states act as energy towers within the low-lying excited
spectrum, indicating that they are quantum many-body scars.
\end{abstract}

\affiliation{School of Physics, Nankai University, Tianjin 300071, China}
\maketitle

\section{Introduction}

One of the foremost roadblocks to quantum simulation and quantum information
tasks is thermalization. Once a system relaxes to equilibrium, all traces of
its initial state are inexorably erased. Yet exceptions are anticipated,
certain carefully prepared, sluggishly thermalizing states may preserve
their imprint far longer than typical. Indeed, it is now well documented
that non-integrable systems can evade thermalization altogether when rare,
non-thermal eigenstates, called quantum many-body scars (QMBS), intervene 
\cite%
{Shiraishi2017,Moudgalya2018,PhysRevB.98.235156,Khemani2019,Ho2019,Shibata2020,McClarty2020,Richter2022,Jeyaretnam2021,turner2018weak,Turner2018,Shiraishi_2019,Lin2019,Choi2019,Khemani2020,Dooley2020,Dooley2021}%
. These non-thermal states are typically embedded within the bulk spectrum
of the system and span a subspace in which initial states fail to thermalize
and instead exhibit periodic behavior. The central goal of this field is to
identify quantum scars in a broad range of non-integrable many-body systems.
On the other hand, a contemporary challenge in condensed-matter physics is
to search for long-lived, non-thermal excited states exhibiting macroscopic,
long-range order. These states are expected to serve as a valuable resource
for both quantum simulation and quantum information processing.

A growing body of models has recently been shown to host QMBS, prompting
attempts to subsume them within unified, systematic frameworks \cite%
{Shiraishi2017Systematic,Mark2020Unified,Moudgalya2020ensure,Pakrouski2020Many,Ren2021Quasisymmetry,Nicholas2020From}%
. Among them, the restricted spectrum generating algebra (RSGA) formalism
introduced in Ref. \cite{Moudgalya2020ensure} provides a classification of
QMBS that lies at the focus of this work. It reveals the features and
structure of a class of Hamiltonians that possess an energy tower exactly.
However, such rigorous rules will inevitably overlook many candidate systems
that possess an approximate energy tower. Little is known about how the
energy tower forms in situations where the RSGA conditions are not exactly
satisfied.

In this work, we focus on a system that approximately obeys the RSGA. We
study the ferromagnetic XXZ Heisenberg model on a Lieb lattice. When a
resonant staggered magnetic field is applied, an exact set of eigenstates
can be constructed within the RSGA framework. Notably, we find that many
other eigenstates can be approximately constructed in the same manner. We
refer to this extended formalism as the quasi-RSGA. Numerical simulations
show that these states act as energy towers within the low-lying excited
spectrum, indicating that they are quantum many-body scars.

Our finding enhances the feasibility of observing the quantum scar in
experiment. Atomic system is an excellent test-bed for quantum simulator in
experiments \cite%
{zhang2017observation,bernien2017probing,barends2015digital,davis2020protecting,signoles2021glassy,trotzky2008time,gross2017quantum}%
, stimulating theoretical studies on the dynamics of quantum spin systems.
These studies not only capture the properties of many artificial systems,
but also provide tractable theoretical examples for understanding
fundamental concepts in physics. As a paradigmatic quantum spin model, the
Heisenberg XXZ model exhibits strong correlations and its dynamical
properties attract the attention from both condensed matter physics and
mathematical-physics communities \cite%
{Keselman2020,bera2020dispersions,Chauhan2020,Babenko2021}. The Lieb lattice
is a two-dimensional lattice model that has attracted significant attention
in condensed matter physics, particularly due to its unique properties
related to flat bands and topological states.\textbf{\ }Furthermore, recent
experimental advances in cold-atom systems enable realizations of the XXZ
chain and preparation of certain initial states \cite%
{fukuhara2013microscopic,jepsen2020spin}, providing an ideal platform for
studying nonequilibrium quantum dynamics. In this context, the results of
this work can be demonstrated experimentally.

The structure of this paper is as follows. In Sec. \ref{Model Hamiltonian
and RSGA}, we introduce the model Hamiltonian and briefly review the concept
of RSGA. In Sec. \ref{Quasi-RSGA condition}, we present the concept of RSGA
and demonstrate it through the concrete model. In Sec. \ref{Dynamic
demonstrations}, We utilize numerical simulations for the dynamic
demonstration of our conclusions regarding quantum many-body scars. Finally,
in Sec. \ref{Summary}, we provide a summary and discussion.

\section{Model Hamiltonian and RSGA}

\label{Model Hamiltonian and RSGA}

\begin{figure}[t]
\centering
\includegraphics[width=0.45\textwidth]{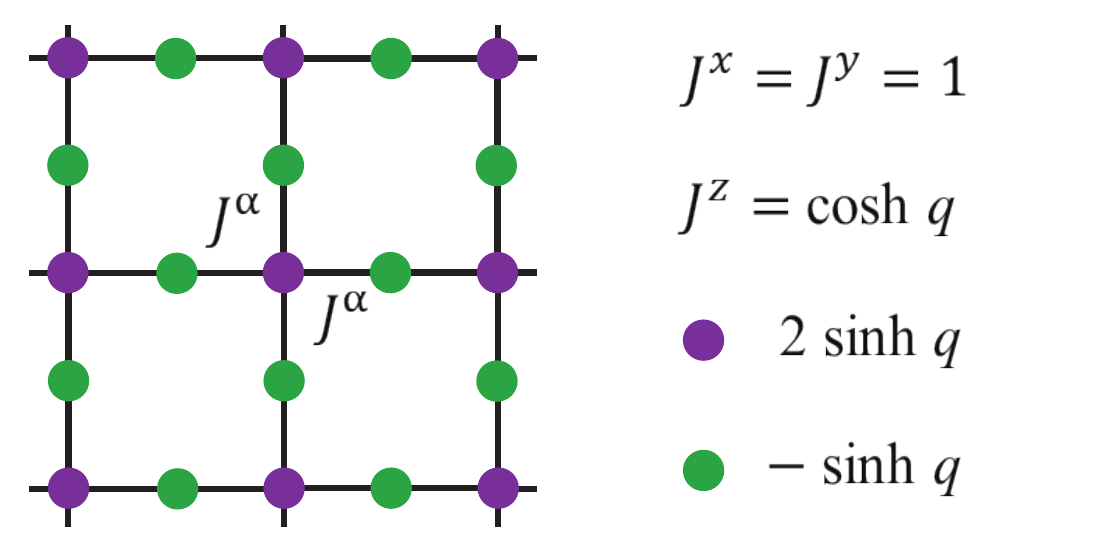}  
\caption{Schematic of a Lieb lattice in which each site is occupied by a $1/2 $ spin. The two sublattices are indicated in purple (corner spins), and
		green (edge spins). The nearest-neighbor couplings $J^{\protect\alpha }$ ($\protect\alpha =x,y,z$) along the horizontal and vertical directions, as
		well as the magnetic fields, are also indicated. For any given $q$, it is
		shown that the ground state of the system can be exactly constructed, while
		a set of excited states can be approximately constructed.} \label{fig1}
\end{figure}

We consider a XXZ Heisenberg quantum spin systems on a two-dimensional Lieb
lattice, which can be viewed as a square lattice with an additional site at
the center of each square unit. When only nearest-neighbor (NN) coupling is
considered, it is bipartite lattice, consisting of sublattice A and B. The
Hamiltonian is written in the form%
\begin{eqnarray}
H &=&-\sum_{\left\langle i,j\right\rangle }\left( \sum_{\alpha
=x,y,z}J^{\alpha }s_{i}^{\alpha }s_{j}^{\alpha }-\frac{J^{z}}{4}\right) 
\notag \\
&&+h(2\sum_{j\in A}s_{j}^{z}-\sum_{j\in B}s_{j}^{z}),  \label{H}
\end{eqnarray}%
where $s_{j}^{\alpha }$\ is the $\alpha $-component spin operator at site $j$%
, $\left\langle i,j\right\rangle $ denotes NN sites, and $\left\{ J^{\alpha
}\right\} $ are the coupling constant between\ nearest neighbors. The
external fields are opposite for two sets of spins in two sublattices. In
this work, we focus on the case with $J^{x}=J^{y}=1$ and $J^{z}=\cosh q$.
The field is taken in the resonant value $h=\sinh q$, for a given real
number $q$. The geometry and thr Hamiltonian of a two-dimensional quantum
spin Lieb lattice with periodic boundary conditions in both two directions
is illustrated in Fig. \ref{fig1}.

To characterize the feature of the Hamiltonian, we introduce a set of
operators

\begin{eqnarray}
\zeta ^{\pm } &=&s_{A}^{\pm }+e^{\pm q}s_{B}^{\pm }, \\
\zeta ^{z} &=&\frac{1}{2}\left[ \zeta ^{+},\zeta ^{-}\right]
=s_{A}^{z}+s_{B}^{z},
\end{eqnarray}%
which satisfy the Lie algebra for any\textbf{\ }$q$. Here, the collective
operators are defined as\textbf{\ }$s_{A}^{\alpha }=\sum_{j\in
A}s_{j}^{\alpha }$\textbf{\ }and\textbf{\ }$s_{B}^{\alpha }=\sum_{j\in
B}s_{j}^{\alpha }$\textbf{\ (}$\alpha =\pm ,z$\textbf{), }respectively.%
\textbf{\ }Note that\textbf{\ }$\zeta ^{+}$\textbf{\ }is not the Hermitian
conjugation of\textbf{\ }$\zeta ^{-}$\textbf{.} In this section, we will
employ the operators $\zeta ^{\pm }$\ to construct the exact eigenstates of
the Hamiltonian, based on the spectrum generating algebra (SGA) \cite%
{Barut1965Dynamical,DOTHAN1965148}. We start with the simplest case with $q=0
$, at which the Hamiltonian reduces the isotropic Heisenberg Hamiltonian
with zero external field. Meanwhile, the operators $\left\{ \zeta ^{\pm
},\zeta ^{z}\right\} $ become total spin operators and the Hamiltonian
possesses SU(2) symmetry, i.e., $\left[ \zeta ^{\pm },H\right] =\left[ \zeta
^{z},H\right] =0$. It allows us to construct a set of eigenstates of $H$ by
employing the SGA. In fact, for a given eigenstate $\left\vert \psi
_{0}\right\rangle $\ of the Hamiltonian $H$\ with energy $E_{0}$, state $%
\left( \zeta ^{\pm }\right) ^{n}\left\vert \psi _{0}\right\rangle $ is also
eigenstates with energy $E_{0}$,\ if $\left( \zeta ^{\pm }\right)
^{n}\left\vert \psi _{0}\right\rangle \neq 0$. Furthermore, state $\left(
\zeta ^{\pm }\right) ^{n}\left\vert \psi _{0}\right\rangle $\ is also an
eigenstate of the operator $\left( \zeta ^{+}\zeta ^{-}+\zeta ^{+}\zeta
^{-}\right) /2+\left( \zeta ^{z}\right) ^{2}$ \cite{Zhang_2023}. In this
sense, such eigenstates are the result of the SU(2) symmetry.

Now, we turn to the case with nonzero $q$. In this situation, the SU(2)
symmetry is broken due the result directly%
\begin{equation}
\left[ \zeta ^{\pm },H\right] =c^{\pm },
\end{equation}%
with 
\begin{eqnarray}
c^{\pm } &=&\pm \sum_{\left\langle i,j\right\rangle ,i\in A}[(\cosh qe^{\pm
q}-1)s_{i}^{z}s_{j}^{\pm }+(\cosh q-e^{\pm q})s_{i}^{\pm }s_{j}^{z}]  \notag
\\
&&\pm \sinh q(e^{\pm q}s_{B}^{\pm }-2s_{A}^{\pm }).
\end{eqnarray}%
It seems that the state $\left( \zeta ^{\pm }\right) ^{n}\left\vert \psi
_{0}\right\rangle $\ is not the eigenstate of $H$. However, straightforward
derivations show that%
\begin{equation}
\left[ \zeta ^{\pm },\left[ \zeta ^{\pm },H\right] \right] =0,
\end{equation}%
and%
\begin{equation}
c^{+}\left\vert \Downarrow \right\rangle =c^{-}\left\vert \Uparrow
\right\rangle =0,
\end{equation}%
where states $\left\vert \Downarrow \right\rangle $ and $\left\vert \Uparrow
\right\rangle $\ are the situated ferromagnetic states, satifying $%
s_{j}^{z}\left\vert \Downarrow \right\rangle =-\frac{1}{2}\left\vert
\Downarrow \right\rangle $ and $s_{j}^{z}\left\vert \Uparrow \right\rangle =%
\frac{1}{2}\left\vert \Uparrow \right\rangle $\ for any $j$. According to
the RSGA, the states $\left( \zeta ^{+}\right) ^{n}\left\vert \Downarrow
\right\rangle $ and $\left( \zeta ^{-}\right) ^{n}\left\vert \Uparrow
\right\rangle $\ ($n\in \left[ 0,N\right] $) are the degenerate eigenstates
of\textbf{\ }$H$\textbf{\ }with zero energy. In this context, we take $%
\left\vert \psi _{0}\right\rangle =\left\vert \Downarrow \right\rangle $%
\textbf{\ }or\textbf{\ }$\left\vert \Uparrow \right\rangle $, both with%
\textbf{\ }$E_{0}=0$\textbf{. }It is straightforward to show that these
states are the ground states of the Hamiltonian $H$\textbf{.} Indeed, the
Hamiltonian given in Eq. (\ref{H}) is the sum of a set of dimers, $%
H=\sum_{\left\langle i,j\right\rangle }H_{ij}$, where%
\begin{equation}
H_{ij}=-\sum_{\alpha =x,y,z}J^{\alpha }s_{i}^{\alpha }s_{j}^{\alpha }+\frac{%
J^{z}}{4}+h(s_{i}^{z}-s_{j}^{z}).
\end{equation}%
We note that the groundstate energy of each dimer $H_{ij}$\ is zero, which
confirms our conclusion. However, we wish to emphasize that the two
eigenstates $\left( \zeta ^{+}\right) ^{n}\left\vert \Downarrow
\right\rangle $\ and $\left( \zeta ^{-}\right) ^{N-n}\left\vert \Uparrow
\right\rangle $\ are identical.

In parallel, there is another simple model that shares the same physics as
the Lieb lattice model. We consider a Hamiltonian of a two-site dimer, which
reads

\begin{eqnarray}
H_{\text{D}} &=&L_{1}^{x}L_{2}^{x}+L_{1}^{y}L_{2}^{y}+\cosh
q(L_{1}^{z}L_{2}^{z}-\frac{MN}{4})  \notag \\
&&+\sinh q(\frac{N}{2}L_{1}^{z}-\frac{M}{2}L_{2}^{z}),
\end{eqnarray}%
where $L_{1}$ and $L_{2}$ are two angular momentum operators with magnitudes 
$L_{1}=\frac{M}{2}$ and $L_{2}=\frac{N}{2}$, and they follow the commutation
relation of angular-momentum operators. Introducing operator

\begin{equation}
L^{+}=L_{1}^{+}+e^{q}L_{2}^{+},
\end{equation}%
straightforward derivations show that%
\begin{eqnarray}
\left[ L^{+},\left[ L^{+},H_{\text{D}}\right] \right]  &=&0, \\
\left[ L^{+},H_{\text{D}}\right] \left\vert \Downarrow \right\rangle  &=&0,
\\
H_{\text{D}}\left\vert \Downarrow \right\rangle  &=&0,
\end{eqnarray}%
which meet the conditions of the RSGA. Then we conclude that the eigenstates
of $H_{\text{D}}$\ can be expressed in the form%
\begin{equation}
\left\vert \psi _{\text{D}}^{l}\right\rangle
=(L_{1}^{+}+e^{q}L_{2}^{+})^{l}\left\vert \Downarrow \right\rangle .
\end{equation}%
We note that when we take $L_{1}^{+}=s_{A}^{+}$ and $L_{2}^{+}=s_{B}^{+}$,
the two Hamiltonians $H$\ and $H_{\text{D}}$\ share the same set of
eigenstates. This equivalence provides a clear physical picture of the
ground states of the quantum spin Lieb lattice. The same result can also be
obtained for the operator $L^{-}=L_{1}^{-}+e^{-q}L_{2}^{-}$. 

In addition, when a uniform magnetic field is added, i.e., $H\rightarrow
H+\mu \left( s_{A}^{z}+s_{B}^{z}\right) $\ or $H_{\text{D}}\rightarrow H_{%
\text{D}}+\mu \left( L_{1}^{z}+L_{2}^{z}\right) $, these degenerate states
are left. The corresponding energy levels are\ equally spaced. They act as
energy towers within the low-lying excited spectrum, indicating that they
are quantum many-body scars.

\section{Quasi-RSGA condition}

\label{Quasi-RSGA condition}

In the last section, we have shown that state $\left( \zeta ^{\pm }\right)
^{n}\left\vert \psi _{0}\right\rangle $\ is an eigenstate of $H$, if we take 
$\left\vert \psi _{0}\right\rangle =\left\vert \Downarrow \right\rangle \ $%
or $\left\vert \Uparrow \right\rangle $. A natural question is whether there
exist other states $\left\vert \psi _{0}\right\rangle $ that meet the RSGA
condition. However, it is hard to find such a state $\left\vert \psi
_{0}\right\rangle $ beyond $\left\vert \Downarrow \right\rangle \ $and $%
\left\vert \Uparrow \right\rangle $, since obtaining other exact eigenstates
of $H$ itself is a challenge. For a finite system, one can search for such
eigenstates by numerical simulations.

Our strategy is as follows: (i) All the eigenstates $\left\{ \left\vert \phi
_{n}^{m}\right\rangle \right\} $\ with energy $\left\{ E_{n}^{m}\right\} $\
in each invariant subspace indexed by $m$ are obtained by numerical
diagonalization. (ii) One can construct a set of states $\left\{ \left\vert
\varphi _{n}^{m+1}\right\rangle \right\} $ by the mapping $\left\vert
\varphi _{n}^{m+1}\right\rangle =\zeta ^{+}\left\vert \phi
_{n}^{m}\right\rangle /\left\vert \zeta ^{+}\left\vert \phi
_{n}^{m}\right\rangle \right\vert $ (or $\zeta ^{-}\left\vert \phi
_{n}^{m}\right\rangle $). (iii) Determine whether a state $\left\vert
\varphi _{n}^{m+1}\right\rangle $\ is an eigenstate of $H$ or not. To
estimate the distance between a state $\left\vert \varphi
_{n}^{m+1}\right\rangle $\ and an eigenstate, one can calculate the variance
of the operator $H$ in the state $\left\vert \varphi _{n}^{m+1}\right\rangle 
$:

\begin{equation}
\Delta H^{2}=\left\langle \varphi _{n}^{m+1}\right\vert H^{2}\left\vert
\varphi _{n}^{m+1}\right\rangle -\left\langle \varphi _{n}^{m+1}\right\vert
H\left\vert \varphi _{n}^{m+1}\right\rangle ^{2}.  \label{variance}
\end{equation}%
If $\left\vert \varphi _{n}^{m+1}\right\rangle $ is an eigenstate, the
variance $\Delta H^{2}$ should be zero. A small variance indicates that the
state is an approximate eigenstate.

In practice, it is rare to have an exact eigenstate in the form $\left\vert
\varphi _{n}^{m+1}\right\rangle $, and constructing a set of approximate
eigenstates is also useful.\ We are interested in the case, where one of the
RSGA conditions is satisfied in a approximation manner, i.e.,%
\begin{eqnarray}
\left[ \zeta ^{+},H\right] \left\vert \phi _{n}^{m}\right\rangle &\approx &0,
\\
\text{or }H\zeta ^{+}\left\vert \phi _{n}^{m}\right\rangle &\approx
&E_{n}^{m}\zeta ^{+}\left\vert \phi _{n}^{m}\right\rangle .
\end{eqnarray}

\begin{figure*}[tbph]
\centering
\includegraphics[width=1\textwidth]{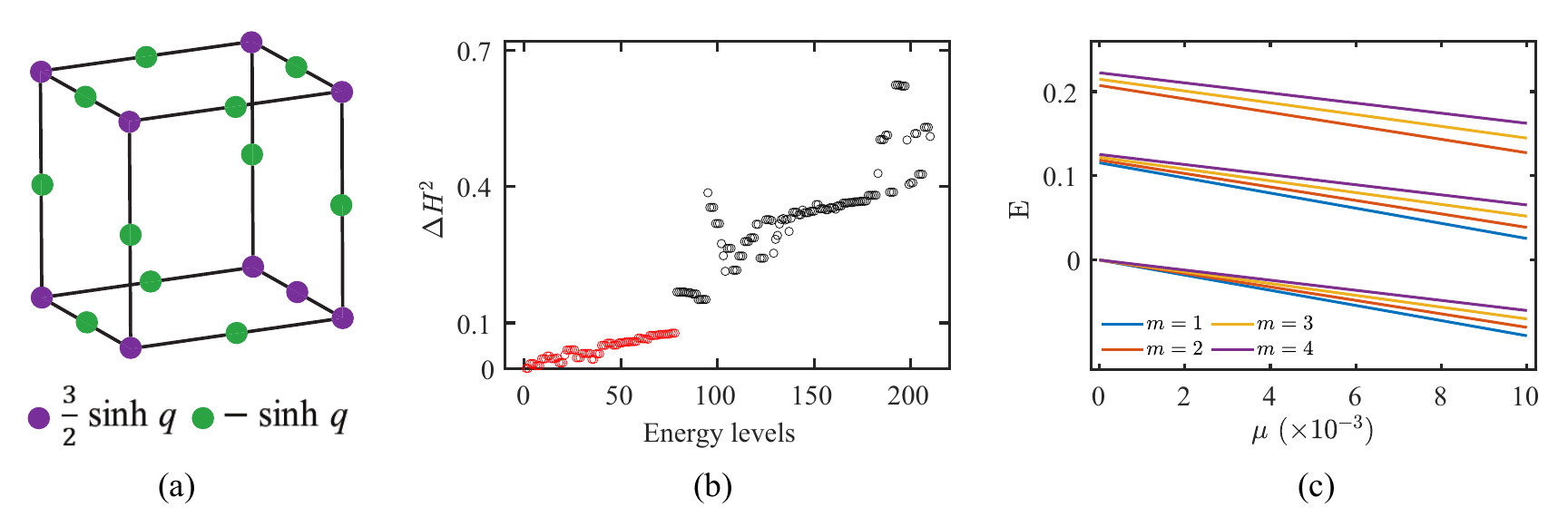}
\caption{(a) Schematic of a Lieb lattice with PBC in both directions with $%
N=20$. The system parameters are the same as that in Fig. \protect\ref{fig1}
. (b) Plots of the variance defined in Eq. (\protect\ref{variance}) for
selected low-lying eigenstates with small $m$ of the Hamiltonian with $N=20$%
. As expected, we see that the variances for the ground state are zero, and
there exist some excited eigenstates (denoted by red empty circles) with
very small variances. (c) Plots of the eigenstate energies associated with
the states $\left\vert \protect\phi _{n}^{m}\right\rangle $ with small
derived variance. The energy levels resemble the Zeeman effect, with
perfectly and nearly uniformly spaced multiplet splitting of the spectral
lines corresponding to the ground states and excited states, respectively.
These sets of eigenstates form energy towers that act as quantum scars.}
\label{fig2}
\end{figure*}

This allows the construction of a set of approximate eigenstates and is
referred to as the quasi-RSGA condition. To demonstrate this point,
numerical simulations are performed for the finite size quantum spin Lieb
lattice. We consider a cluster with PBC in both two directions, which is
illustrated in Fig. \ref{fig2}. It is the minimal Lieb lattice with PBCs in
both directions. However, it remains a challenge to diagonalize a spin
system with $N=20$ in the full Hilbert space. Here, we perform the numerical
simulation in each invariant subspace indexed by $m$. In Fig. \ref{fig2}, we
plot the variance $\Delta H^{2}$ for the low-lying states $\left\vert \phi
_{n}^{m}\right\rangle $\ with small $m$. As expected, we have following
observations. (i) The variances are zero for the ground states. (ii) There
indeed exist some energy levels with small variance, meeting the quasi-RSGA
condition. (iii) Such energy levels are quasi-degenerate. When a uniform
field is applied, the Hamiltonian becomes $H\rightarrow $ $H+$ $\mu
(s_{A}^{z}$ $+s_{B}^{z})$. The eigenstates remain unchanged due to the fact
that $\left[ H,s_{A}^{z}+s_{B}^{z}\right] =0$. As shown in Fig. \ref{fig2},
the corresponding energy levels become quasi equal-spaced levels. These
results indicate that, besides the ground states, there exist many set of
energy levels with nearly uniformly spaced splitting. These sets of
eigenstates form energy towers that act as quantum scars.

\section{Dynamic demonstrations}

\label{Dynamic demonstrations}

In this section, we turn to the dynamical demonstration of our results. We
will focus on the system with OBC due to the following two concerns. First,
we will show that our results hold true for the system with OBC. Second, a
sample with OBC can be a smaller system, which allows us to perform
numerical simulations in the full Hilbert space. Specifically, we consider a
system with $N=13$. In order to meet the RSGA conditions for the
groundstates of the system, the fields on the boundary spins should be
modified. Unlike the lattice systems presented in Fig. \ref{fig1} and \ref%
{fig2}, where the fields for sublattices A and B are uniform, respectively,
the fields for lattice A are not uniform due to the open boundary conditions.

In this situation, we classify the lattice into three sublattices A, B and
C. The corresponding Hamiltonian $H_{\text{OBC}}$ is illustrated in Fig. \ref%
{fig3}(a). Accordingly, the operator $\zeta ^{\pm }$\ is given by%
\begin{equation}
\zeta ^{\pm }=s_{A}^{\pm }+e^{\pm q}s_{B}^{\pm }+s_{C}^{\pm }.
\end{equation}%
The method for constructing the Hamiltonian $H_{\text{OBC}}$\ is detailed in
the reference \cite{LSJY2025Condensate}. Notably, two operators $H_{\text{OBC%
}}$\ and $\zeta ^{+}$\ satisfy the following RSGA conditions

\begin{eqnarray}
H_{\text{OBC}}\left\vert \Downarrow \right\rangle &=&0, \\
\left[ \zeta ^{+},H_{\text{OBC}}\right] \left\vert \Downarrow \right\rangle
&=&0, \\
\left[ \zeta ^{+},\left[ \zeta ^{+},H_{\text{OBC}}\right] \right] &=&0,
\end{eqnarray}%
which can be shown by directly derivation. In parallel, the corresponding
relations for $\zeta ^{-}$\ can also be obtained. Accordingly, we can
construct a set of degenerate ground states, given by

\begin{equation}
\left\vert \psi _{\text{g}}^{l}\right\rangle =\frac{1}{\sqrt{\Omega _{\text{g%
}}^{l}}}\left( \zeta ^{+}\right) ^{l}\left\vert \Downarrow \right\rangle ,
\end{equation}%
which satisfy%
\begin{equation}
H_{\text{OBC}}\left\vert \psi _{\text{g}}^{l}\right\rangle =0,
\end{equation}%
with $\Omega _{\text{g}}^{l}$ being the normalization factor and$\ l\in
\lbrack 0,13]$. Furthermore, we also find that the first excited state $%
\left\vert \text{e}\right\rangle $ can be obtained exactly. The exact form
of the state $\left\vert \text{e}\right\rangle $\ is depicted in Fig. \ref%
{fig3}(a), which satisfies the Schrodinger equation%
\begin{equation}
H_{\text{OBC}}\left\vert \text{e}\right\rangle =(\cos q-\sqrt{\cosh ^{2}q-1/4%
})\left\vert \text{e}\right\rangle .
\end{equation}

\begin{figure*}[t]
\centering
\includegraphics[width=1\textwidth]{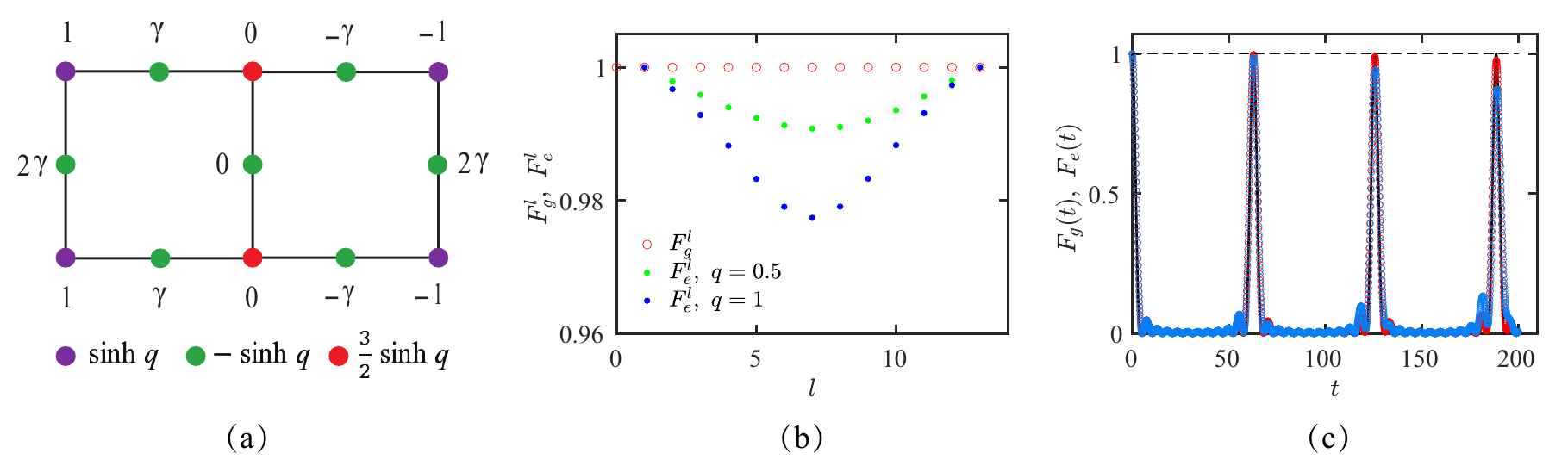}
\caption{(a) Schematic of a Lieb lattice with OBC in both directions with $%
N=13$. The system parameters are indicated explicitly. The coupling
constants are the same as that in Fig. \protect\ref{fig1} and \protect\ref%
{fig2}, while the on-site fields are different. The factors indicated at
each sites denote the amplitudes of the first-excited state with $m=1$,
where $\protect\gamma =(2\sinh q$ $+\protect\sqrt{\cosh ^{2}q-1})/3$.\ (b)
Plots of the square mode of overlaps for the ground states and first excited
states, given by Eq. (\protect\ref{Fe}), obtained by exact diagonalization
for the cases with different values of $q$. It indicates that the states
constructed based on the ground state are exact eigenstates, while those
based on the first excited state are nearly exact eigenstates. A smaller
value of $q$ leads to a better approximation. (c) Plots of the fidelities,
given by Eqs. (\protect\ref{Fg(t)}) and (\protect\ref{Fe(t)}), for two
initial states given by Eqs. (\protect\ref{PHIg}) and (\protect\ref{PHIe})
for the cases with different values of $q$. Here, $F_{\text{g}}^{l}$ is
plotted as a black line, while $F_{\text{e}}^{l}$\ is plotted as red empty
circles for the case with $q=0.5$ and as blue empty circles for the case
with $q=1.0$. We can see that the time evolution of the first type of
initial state $\left\vert \Phi _{\text{g}}(0)\right\rangle $ exhibits
perfect periodic revival, while the second type of initial state $\left\vert
\Phi _{\text{e}}(0)\right\rangle $\ exhibits near-perfect periodic revival.
For the state $\left\vert \Phi _{\text{e}}(0)\right\rangle $, lower values
of $q$ yield better revival.}
\label{fig3}
\end{figure*}
The variances of the operator $H_{\text{OBC}}$ for the normalized state $%
\frac{1}{\sqrt{4(1+3\gamma ^{2})}}\zeta ^{+}\left\vert \text{e}\right\rangle 
$\ are $0.0051$ and $0.0141$, for $q=0.5$ and $q=1.0$, respectively. This
indicates that a set of states, given by 
\begin{equation}
\left\vert \psi _{\text{e}}^{l}\right\rangle =\frac{1}{\sqrt{\Omega _{\text{e%
}}^{l}}}\left( \zeta ^{+}\right) ^{l-1}\left\vert \text{e}\right\rangle ,
\end{equation}%
with $\Omega _{\text{e}}^{l}$ is the normalization factor and $l\in \lbrack
1,13]$, are quasi eigenstates of the Hamiltonian $H_{\text{OBC}}$. On the
other hand, the ground states $\left\{ \left\vert \phi _{\text{g}%
}^{l}\right\rangle ,l\in \left[ 0,13\right] \right\} $\ and the first
excited states $\left\{ \left\vert \phi _{\text{e}}^{l}\right\rangle ,l\in %
\left[ 1,13\right] \right\} $ in each invariant subspace indexed by $l$ can
be obtained by numerical simulations. Here, $l$\ can be directly obtained
from the eigenvalue of the conserved observable $%
s_{A}^{z}+s_{B}^{z}+s_{C}^{z}$.\ 

This allows us to verify our predictions by estimating the difference
between the constructed state and the corresponding exact eigenstate. We
employ the fidelity, which is the mode square of the overlap between two
states, to quantify the closeness between them. Obviously, we have%
\begin{equation}
F_{\text{g}}^{l}=\left\vert \left\langle \psi _{\text{g}}^{l}\right.
\left\vert \phi _{\text{g}}^{l}\right\rangle \right\vert ^{2}=1,  \label{Fg}
\end{equation}%
with $l\in \left[ 0,13\right] $ for any value of $q$. Similarly, the
quantities 
\begin{equation}
F_{\text{e}}^{l}=\left\vert \left\langle \psi _{\text{e}}^{l}\right.
\left\vert \phi _{\text{e}}^{l}\right\rangle \right\vert ^{2},  \label{Fe}
\end{equation}%
can be obtained by numerical simulation with $l\in \left[ 1,13\right] $ for
a given value of $q$. We plot the quantities $F_{\text{g}}^{l}$\ and $F_{%
\text{e}}^{l}$\ for two representative values of $q$\ in the Fig. \ref{fig3}%
(b), which indicate that the set of states $\left\{ \left\vert \psi _{\text{e%
}}^{l}\right\rangle \right\} $\ are nearly exact eigenstates of $H_{\text{OBC%
}}$ as we predicted. Numerical simulations show that there exist several
other sets of excited states with similar features. Here, we only focus on
the ground and first excited states. The implications of these results are
obvious. When applying a uniform external field by taking $H_{\text{OBC}%
}\rightarrow $ $H_{\text{OBC}}$ $+\mu \left(
s_{A}^{z}+s_{B}^{z}+s_{C}^{z}\right) $, the eigenstates $\left\{ \left\vert
\phi _{\text{g}}^{l}\right\rangle \right\} $\ and the first excited states $%
\left\{ \left\vert \phi _{\text{e}}^{l}\right\rangle \right\} $\ become
energy towers as quantum many-body scars. It leads to revivals of initial
states primarily supported on the scar subspace, but fail to reproduce
thermal expectation values of local observables in system.

To demonstrate and verify the conclusion, numerical simulations are
performed for the time evolution of the two types of initial states in the
form%
\begin{equation}
\left\vert \Phi _{\text{g}}(0)\right\rangle =\frac{1}{\sqrt{14}}%
\sum_{l=0}^{13}\left\vert \phi _{\text{g}}^{l}\right\rangle ,  \label{PHIg}
\end{equation}%
and%
\begin{equation}
\left\vert \Phi _{\text{e}}(0)\right\rangle =\frac{1}{\sqrt{13}}%
\sum_{l=1}^{13}\left\vert \phi _{\text{e}}^{l}\right\rangle ,  \label{PHIe}
\end{equation}%
which are simply equally superpositions of the two sets of exact eigenstates 
$\left\{ \left\vert \phi _{\text{g}}^{l}\right\rangle \right\} $\ and $%
\left\{ \left\vert \phi _{\text{e}}^{l}\right\rangle \right\} $,
respectively.

Based on our results, the dynamics in the quantum scars depends on the
initial state and the parameter $q$, and can be characterized by the
fidelities%
\begin{equation}
F_{\text{g}}(t)=\left\vert \left\langle \Phi _{\text{g}}(0)\right\vert
e^{-iH_{\text{OBC}}t}\left\vert \Phi _{\text{g}}(0)\right\rangle \right\vert
^{2},  \label{Fg(t)}
\end{equation}%
and%
\begin{equation}
F_{\text{e}}(t)=\left\vert \left\langle \Phi _{\text{e}}(0)\right\vert
e^{-iH_{\text{OBC}}t}\left\vert \Phi _{\text{e}}(0)\right\rangle \right\vert
^{2}.  \label{Fe(t)}
\end{equation}%
We have the following predictions.

(i) In the case with zero $q$, the Hamiltonian $H_{\text{OBC}}$\ has SU(2)
symmetry, and the operator $\zeta ^{\pm }$\ reduces to spin operator. All
the eigenstates $\left\{ \left\vert \phi _{\text{g}}^{l}\right\rangle
\right\} $\ and $\left\{ \left\vert \phi _{\text{e}}^{l}\right\rangle
\right\} $\ can be generated by the spin operator $\zeta ^{\pm }$, and thus
possess equally spaced energy levels. Consequently, the evolved states
exhibit perfect periodic behavior with frequency $\mu $. Both $F_{\text{g}%
}(t)$\ and $F_{\text{e}}(t)$\ exhibit perfect revival at the instants $%
t=2n\pi /\mu $\ (where $n=1,2,3,...$).

(ii) In the case with non-zero $q$, the Hamiltonian $H_{\text{OBC}}$\ does
not have SU(2) symmetry, and the operator $\zeta ^{\pm }$\ is not a spin
operator. However, the eigenstates $\left\{ \left\vert \phi _{\text{g}%
}^{l}\right\rangle \right\} $ are identical to $\left\{ \left\vert \psi _{%
\text{g}}^{l}\right\rangle \right\} $. Consequently, $F_{\text{g}}(t)$\
exhibits perfect revival at the instants $t=2n\pi /\mu $\ (where $%
n=1,2,3,... $), while $F_{\text{e}}(t)$\ exhibits nearly perfect revival\
around the instants $t=2n\pi /\mu $.

Numerical simulations are conducted for the time evolutions to verify our
predictions and to assess the efficiency of the approximation. The numerical
results for the quantities $F_{\text{g}}(t)$\ and $F_{\text{e}}(t)$\ are
plotted in Fig. \ref{fig3}(c). As can be seen in the figure, the quantity $%
F_{\text{g}}(t)$ shows perfect periodic revivals for a given $q$, while the
quantity $F_{\text{e}}(t)$ shows quasi-periodic revivals with relatively
slow decay for a given $q$. A smaller value of $q$ leads to a higher
fidelities.

\section*{Summary}

\label{Summary} In summary, we have extended the RSGA into a quasi-RSGA for
constructing approximate degenerate eigenstates of a ferromagnetic XXZ
Heisenberg model on a Lieb lattice with a resonant staggered magnetic field.
When a uniform external field is applied, each set of approximate
eigenstates becomes energy towers, acting as quantum many-body scars. To
assess the efficiency of these approximate energy towers as quantum scars in
such a spin system, numerical simulations have been performed for finite
systems. We found that the obtained energy towers support near-perfect
revivals under a wide range of system parameters. Our finding provides an
alternative method for constructing quantum many-body scars\ through
resonant external field.

\section*{Acknowledgment}

We acknowledge the support of NSFC (Grants No. 12374461).

\bibliographystyle{unsrt}
\bibliography{reference}

\end{document}